\documentclass[preprint,3p,twocolumn]{elsarticle_arxiv}

\usepackage{amssymb}
\usepackage{amsthm}
\usepackage{amsmath}
\usepackage{graphicx}
\usepackage{caption}
\usepackage{subcaption}
\usepackage{xcolor}
\usepackage{bm}
\usepackage{xurl}
\usepackage{caption}
\usepackage{float}
\usepackage[colorlinks=true, allcolors=blue]{hyperref}
\usepackage{multirow}
\usepackage{array,makecell}
\usepackage[normalem]{ulem}

\begin{document}

\begin{frontmatter}

\title{On the improved performances of FLUKA~\mbox{v4-4.0} \\ in out-of-field proton dosimetry}

\author[inst1,inst2,inst3]{Alexandra-Gabriela \c{S}erban}

\affiliation[inst1]{organization={European Organization for Nuclear Research},
    addressline={Esplanade des Particules 1},
    city={1211 Geneva 23},
    country={Switzerland}}

\affiliation[inst2]{organization={Faculty of Physics, University of Bucharest},
    addressline={405 Atomistilor},
    city={077125 Bucharest-Magurele},
    country={Romania}}
    
\affiliation[inst3]{organization={“Horia Hulubei” National Institute of Physics and Nuclear Engineering},
    addressline={30 Reactorului},
    city={\\ 077125 Bucharest-Magurele},
    country={Romania}}

\affiliation[inst4]{organization={Departamento de Física Atómica, Molecular y Nuclear, Universidad de Granada},
    addressline={E-18071},
    city={Granada},
    country={Spain}}

\affiliation[inst5]{organization={Instituto de Investigación Biosanitaria (ibs.GRANADA), \\ Complejo Hospitalario Universitario 
de Granada/Universidad de Granada},
    addressline={E-18016},
    city={Granada},
    country={Spain}}

\author[inst4,inst5]{Juan Alejandro de la Torre González} 

\author[inst4,inst5]{\\ Marta Anguiano}

\author[inst4,inst5]{Antonio M. Lallena}

\author[inst1]{Francesc {Salvat-Pujol}\corref{cor1}}
\ead{fluka.team@cern.ch}
\cortext[cor1]{Corresponding author}

\begin{abstract}

    A new model for the nuclear elastic scattering of protons below $250$~MeV has been recently included in 
    FLUKA~\mbox{v4-4.0}, motivated by the evaluation of radiation effects in electronics. Nonetheless, proton nuclear 
    elastic scattering plays a significant role also in proton dosimetry applications, for which the new model necessitated 
    an explicit validation. Therefore, in this work a benchmark has been carried out against a recent measurement of 
    radial-depth maps of absorbed dose in a water phantom under irradiation with protons of $100$~MeV, $160~$MeV, and 
    $225$~MeV. Two FLUKA versions have been employed to simulate these dose maps:~\mbox{v4-3.4}, relying on a legacy model 
    for proton nuclear elastic scattering, and~\mbox{v4-4.0}, relying on the new model. The enhanced agreement with 
    experimental absorbed doses obtained with FLUKA~\mbox{v4-4.0} is discussed, and the role played by proton nuclear 
    elastic scattering, among other interaction mechanisms, in various regions of the radial-depth dose map is elucidated. 
    Finally, the benchmark reported in this work is sensitive enough to showcase the importance of accurately characterizing 
    beam parameters and the scattering geometry for Monte Carlo simulation purposes.

\end{abstract}


\end{frontmatter}

\section{Introduction}
\label{sec:intro}

FLUKA~\cite{flukaweb,batt,frontiers} is a general-purpose code for the Monte Carlo (MC) simulation of coupled hadronic and 
electromagnetic radiation showers in complex geometries. It can transport over $60$~particles species with energies from the 
keV to the PeV domain, with neutrons exceptionally tracked down to $0.01$~meV. Naturally, FLUKA is routinely employed in a 
wide variety of applications, ranging from accelerator design and operation~\cite{potoine}, to associated radiation 
protection studies~\cite{bozzato,cables}, the assessment of radiation effects in 
electronics~\cite{lerner,lerner_clear,cecchetto}, and medical applications~\cite{ambient_dose,beamline,brachytherapy}, to 
name but a few.

Recent studies assessing the production of single-event-upsets in electronic devices under proton irradiation revealed 
a significant FLUKA underestimation for proton energies between $1$~and~$10$~MeV~\cite{coronetti}. This shortcoming was 
attributed to a too simplistic model for the nuclear elastic scattering of protons up to FLUKA 
version~\mbox{v4-3.4}~\cite{ranft} (included). In FLUKA~\mbox{v4-4.0} (released on Feb.~14, 2024), a new model for the 
nuclear elastic scattering of protons from Coulomb barrier up to $250$~MeV was included~\cite{serban}, relying on a 
partial-wave analysis of experimental angular distributions~\cite{exfor,zerkin}. This model significantly improved 
FLUKA's capability to evaluate single-event-upset production in electronic devices under irradiation by protons with 
energies from $1$~to~$100$~MeV~\cite{serban,serban_seu}.

Nuclear elastic scattering of protons plays a significant role, not only for the assessment of radiation effects in 
electronics~\cite{serban,akkerman,caron}, but also in proton dosimetry~\cite{mott}. Indeed, proton nuclear elastic 
scattering contributes to the angular spread of proton showers in matter, and therefore has a direct effect on depth-dose 
maps, especially out-of-field (and thus on the dose delivered to nearby healthy tissue). A wide range of benchmarks was 
carried out prior to the public release of FLUKA~\mbox{v4-4.0} to validate its new model for proton nuclear elastic 
scattering~\cite{serban}. Among them, a dedicated FLUKA simulation of a recent measurement of radial-depth ($r$-$z$) maps of 
dose absorbed in a water phantom under proton irradiation~\cite{verbeek} was performed.

The aim of this work, done on behalf of the FLUKA.CERN Collaboration, is twofold: on the one hand, to showcase the improved 
performances of FLUKA~\mbox{v4-4.0} with respect to~\mbox{v4-3.4} in capturing features across various regions of the 
aforementioned experimental $r$-$z$ dose maps and, on the other hand, to emphasize the importance of accurately 
characterizing the radiation source and the scattering geometry for MC simulation purposes. In Section~\ref{sec:exp}, the 
experimental setup used to measure the aforementioned $r$-$z$ dose maps~\cite{verbeek} is summarized. Next, in 
Section~\ref{sec:source}, the proton source adopted in the corresponding FLUKA simulations is explicitly detailed and the 
rest of the simulation setup (geometry and scoring) is documented. In Section~\ref{sec:simulations_noair}, the agreement 
between experimental $r$-$z$ dose maps and those simulated with FLUKA~\mbox{v4-3.4} and~\mbox{v4-4.0} is discussed, putting 
particular emphasis on the significant role played by proton nuclear elastic scattering in various regions of the $r$-$z$ 
dose maps. In Section~\ref{sec:move_source}, the virtues and shortcomings of the Fermi-Eyges proton beam definition adopted 
in~\cite{verbeek} are outlined, and a procedure is proposed to capture features of the angular distribution beyond the 
Gaussian core that the Fermi-Eyges theory focuses on. Finally, in Section~\ref{sec:conclusions}, a summary and conclusions 
are provided. Additionally, a general overview of the role played by proton nuclear elastic scattering and other relevant 
interaction mechanisms in the $r$-$z$ maps of absorbed dose in water under proton irradiation is provided in 
the~\hyperref[app:2D_maps]{Appendix}.

\section{Experimental radial-depth dose maps}
\label{sec:exp}

In a recent work~\cite{verbeek}, detailed $r$-$z$ maps of dose absorbed in a water phantom under irradiation with $100$~MeV, 
$160$~MeV, and $225$~MeV protons were reported. In this section, a summary of the experimental setup aspects necessary for 
the FLUKA simulations below is provided; further details are deferred to the original reference~\cite{verbeek}.

The aforementioned $r$-$z$ maps of absorbed dose were measured with a two-dimensional array of $1020$ ionization 
chambers arranged in a square grid and placed in a water phantom of $40$~cm length. The diameter of each cylindrical 
chamber was $0.42$~cm and the center of the considered proton beam coincided with the center of one ionization chamber, 
as further detailed in~\cite{verbeek}. The reported uncertainties at each experimental point were $2.5$\% in the 
absorbed dose, $\pm 0.07$~cm in depth, and $\pm 0.01$~cm in the radial position. Moreover, the experimental $r$-$z$ dose 
maps were not given in absolute units: they were normalized to the dose at $3$~cm on the beam axis.

N.B.: For the $100$~MeV proton beam, at the radial distance of $2.29$~cm, the labels on the $y$ axis are inaccurate in the 
original reference. Correct experimental absorbed doses have been provided directly by the authors of the original 
paper~\cite{verbeek}.

\section{FLUKA simulation setup}
\label{sec:source}

In~\cite{verbeek}, the spatial profile and angular distribution of the proton beam at the entrance of 
the water phantom are characterized relying on the Fermi-Eyges theory~\cite{eyges,Rossi,GottschalkFE}. In this approach, the 
probability density to find a proton at a position $\boldsymbol{r}$ after a path length $s$ is parametrized as:
\begin{equation}
\label{eq:fermi_eyges}
   \Phi(s;\boldsymbol{r},\theta_x,\theta_y) = F(z;x,\theta_x) F(z;y,\theta_y),
\end{equation}
where
\begin{equation}
\begin{aligned}
\label{eq:F}
   & F(z;x,\theta_x)
   =
   \frac{1}{4 \pi \sqrt{B(z)}} \\
   & \times \exp \left( - \frac{A_0(z) x^2 - 2 A_1(z) x \theta_x + A_2(z) \theta^2_x}{4 B(z)} \right)
\end{aligned}
\end{equation}
is the probability density for a proton at depth $z$ to have a transverse position $x$ and an angle $\theta_x$ between the
direction of motion projected on the $xz$ plane and the $z$ axis; similarly for the projections on the $yz$
plane~\cite{Rossi}. The beam parameters are the variance of the angular distribution ($2A_0$), the variance of the spatial
distribution ($2A_2$), and their covariance ($2A_1$). The emittance of the proton beam is evaluated as $B = A_0 A_2 - A_1^2$.
Transverse positions and angles are sampled following~\cite{verbeek} as:
\begin{equation}
\begin{aligned}
\label{eq:x_theta}
   x        & = \mu_x + \sqrt{2A_2} \xi_1 \\
   \theta_x & = \mu_{\theta_x} + \sqrt{\frac{2A^2_1}{A_2}} \xi_1 + \sqrt{\frac{2B}{A_2}} \xi_2,
\end{aligned}
\end{equation}
where $\mu_x$ is the mean position, $\mu_{\theta_x}$ is the mean angle, and $\xi_{1,2}$ are two normally distributed 
pseudo-random numbers; axial symmetry around the $z$-axis is assumed, therefore the same expressions with the same 
parameters hold for the $y$ components. For completeness, the proton beam parameters at the entrance of the water phantom 
for $100$~MeV, $160$~MeV, and $225$~MeV suggested by~\cite{verbeek}, are reported in Table~\ref{tab:simulation_param}.

\begin{table}[h]
\caption{Fermi-Eyges proton beam parameters for $100$~MeV, $160$~MeV, and $225$~MeV at the entrance of the water 
phantom~\cite{verbeek}.}
\label{tab:simulation_param}
\begin{center}
\begin{tabular}{|l|c|c|c|}
\hline & \makecell{\textbf{100} \\ \textbf{MeV}} & \makecell{\textbf{160} \\ \textbf{MeV}} & \makecell{\textbf{225} \\ \textbf{MeV}} \\
\hline $E$ (MeV)               & $100.150$ & $160.244$ & $225.142$ \\
\hline $\sigma_E$ (MeV)        & $0.614$   & $0.835$   & $0.513$ \\
\hline $\sqrt{2A_2(0)}$ (cm)   & $0.536$   & $0.334$   & $0.320$ \\
\hline $2A_1(0)$ (cm mrad)     & $1.320$   & $0.809$   & $0.773$  \\
\hline $\sqrt{2A_0(0)}$ (mrad) & $6.01$    & $3.52$    & $3.90$ \\
\hline $\mu_x$ (cm)            & $0$       & $0$       & $0$ \\
\hline $\mu_\theta$ (mrad)     & $0$       & $0$       & $0$ \\ 
\hline
\end{tabular}
\end{center}
\end{table}

\begin{figure}
\centering
\includegraphics[width=\columnwidth]{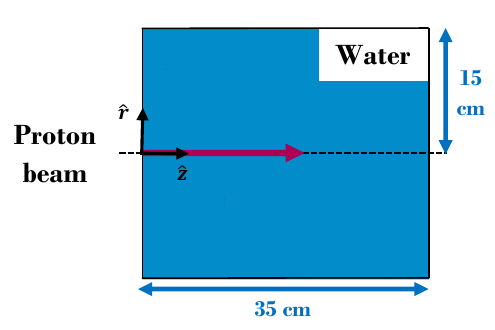}
\caption{Schematic FLUKA simulation geometry showing the proton beam starting at the entrance of the water phantom.}
\label{fig:geom_noair}
\end{figure}

The FLUKA geometry adopted to simulate the $r$-$z$ dose maps of~\cite{verbeek}, set up with the Flair graphical user 
interface~\cite{flair,newflair}, is schematically represented in Fig.~\ref{fig:geom_noair}: a water cylinder of $35$~cm 
length and $15$~cm radius is aligned with the $z$ axis, and a beam of protons is initiated at its entrance, employing the 
Fermi-Eyges theory of Eq.~\eqref{eq:fermi_eyges} to sample the initial position and direction of the primary protons. Three 
beam energies are considered in separate runs: $100$~MeV, $160$~MeV, and $225$~MeV. In all cases, $2.5 \times 10^{8}$ 
primary protons are simulated.

Absorbed dose is scored in the water phantom as a function of depth at various radial distances from the beam axis for the 
three considered energies. For each radial distance, a cylindrical mesh scoring is employed to score the dose absorbed in 
the water phantom. A radial bin width of $0.42$~cm is adopted, corresponding to the diameter of the ionization chamber used 
experimentally~\cite{verbeek}. Table~\ref{tab:scoring} summarizes the radial binning employed for the $100$~MeV, $160$~MeV, 
and $225$~MeV proton beam cases, respectively. Following~\cite{verbeek}, a $1$~mm bin width along the $z$ axis is adopted.

\begin{table}[h]
\caption{Radial binning employed in FLUKA to score the absorbed dose by $100$~MeV, $160$~MeV, and $225$~MeV protons 
in water.}
\label{tab:scoring}
\begin{center}
\begin{tabular}{|c|c|c|c|}
\hline \makecell{$r^{\mathrm{min,max}}_j$ \\ (cm)} & \makecell{\textbf{100} \\ \textbf{MeV}} & \makecell{\textbf{160} \\ 
\textbf{MeV}} & \makecell{\textbf{225} \\ \textbf{MeV}} \\
\hline $r_1^{\mathrm{min}}$ & $0.00$ & $0.00$ & $0.00$  \\
       $r_1^{\mathrm{max}}$ & $0.42$ & $0.42$ & $0.42$  \\
\hline $r_2^{\mathrm{min}}$ & $0.55$ & $1.31$ & $1.31$  \\
       $r_2^{\mathrm{max}}$ & $0.97$ & $1.73$ & $1.73$  \\
\hline $r_3^{\mathrm{min}}$ & $2.08$ & $2.84$ & $2.08$  \\
       $r_3^{\mathrm{max}}$ & $2.50$ & $3.26$ & $2.50$  \\
\hline $r_4^{\mathrm{min}}$ & $2.84$ & $4.36$ & $7.41$  \\
       $r_4^{\mathrm{max}}$ & $3.26$ & $4.78$ & $7.83$  \\
\hline $r_5^{\mathrm{min}}$ & $4.36$ & $5.89$ & $9.69$  \\
       $r_5^{\mathrm{max}}$ & $4.78$ & $6.31$ & $10.11$ \\
\hline $r_6^{\mathrm{min}}$ & $5.89$ & $7.41$ & $10.46$ \\
       $r_6^{\mathrm{max}}$ & $6.31$ & $7.83$ & $10.88$ \\
\hline
\end{tabular}
\end{center}
\end{table}

Electron transport and production cut-offs in FLUKA are set to $50$~keV. This choice, while slightly excessive (the range of 
$50$~keV electrons in water is $\sim$$43$~$\mu$m~\cite{nist,ESTAR}, much shorter than the adopted $r$ and $z$ resolutions), 
ensures that electron histories are stopped when they have practically no chance to contribute to any other $r$-$z$ bin. 
Thus, the scored dose maps are free from particle-range-induced artefacts. For completeness, the production and transport 
thresholds for photons are set to $100$~eV, those for hadrons to $100$~keV, neutrons are transported down to $0.01$~meV, and 
light ions (d, t, $^3$He, and $^4$He) down to $100$~keV; the production and transport thresholds for heavier ions are scaled 
from those of $^4$He by the ratio of mass numbers. Furthermore, the mean excitation energy of water is set to 
I=$78$~eV~\cite{ICRU90}.

\section{FLUKA $r$-$z$ dose maps}
\label{sec:simulations_noair}

\begin{figure*}
\centering
\includegraphics[width=0.95\textwidth]{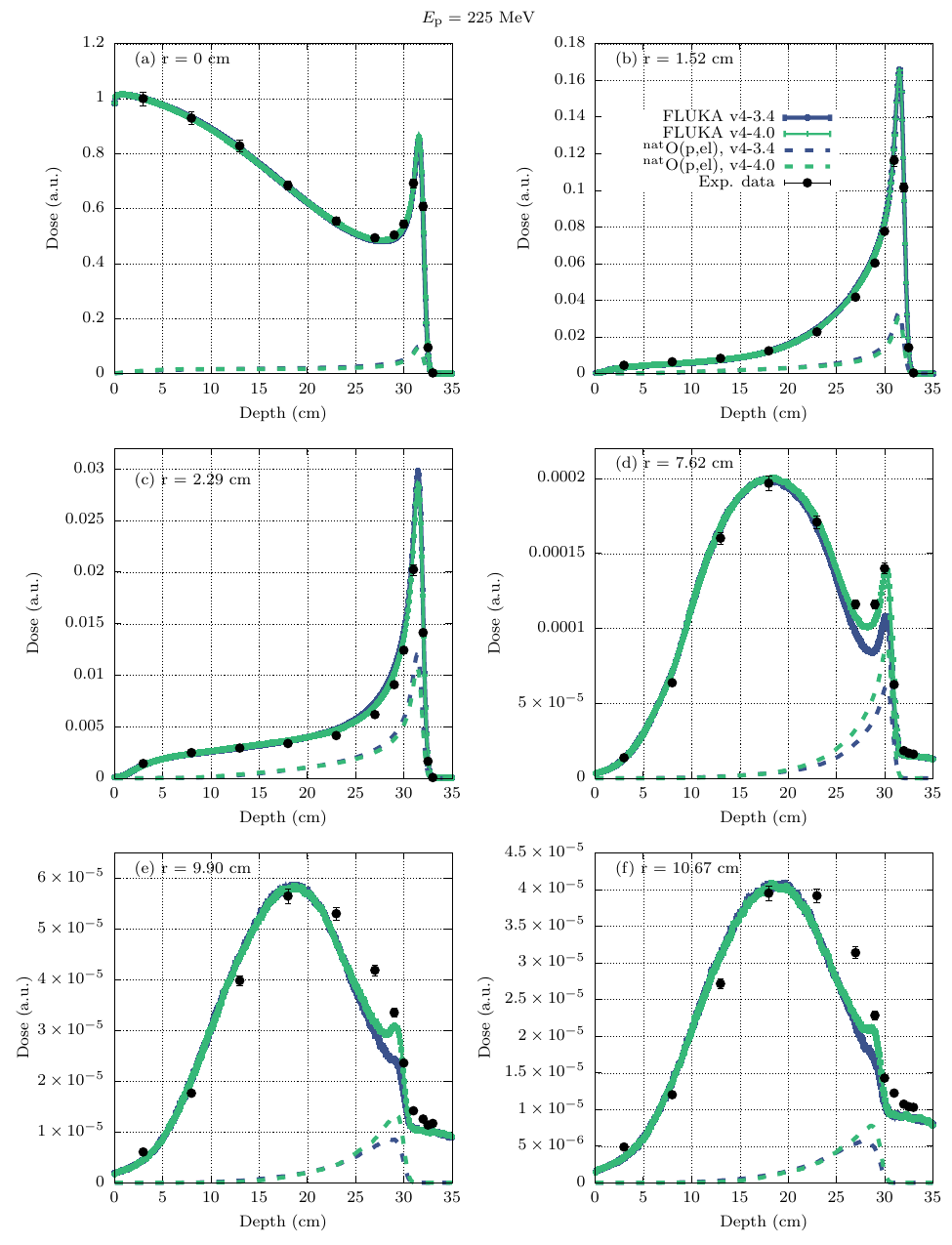}
\caption{Absorbed dose in arbitrary units as a function of depth by $225$~MeV protons in water scored with 
FLUKA~\mbox{v4-3.4} (solid blue) and~\mbox{v4-4.0} (solid green) at the various indicated radial distances. The black 
dots represent experimental absorbed doses~\cite{verbeek}. The blue and green dashed curves show the contribution from 
particle histories where protons underwent a nuclear elastic scattering event on oxygen, scored with 
FLUKA~\mbox{v4-3.4} and~\mbox{v4-4.0}, respectively.}
\label{fig:verbeek_225MeV}
\end{figure*}

Employing the setup described in the foregoing section, the experimental $r$-$z$ dose maps have been simulated with two 
FLUKA versions: \mbox{v4-3.4}, relying on the legacy model for proton nuclear elastic scattering~\cite{ranft}, 
and~\mbox{v4-4.0} including a new model for this interaction mechanism~\cite{serban}. Figure~\ref{fig:verbeek_225MeV} 
displays $r$-$z$ maps of absorbed dose in the water phantom for the $225$~MeV proton beam case. Black dots represent the 
experimental absorbed dose of~\cite{verbeek}, while solid curves result from FLUKA~\mbox{v4-3.4} (blue) and \mbox{v4-4.0} 
(green); dashed curves are discussed below. Dose maps are represented as a function of depth, but not integrated over the 
transverse plane; rather, the radial distance indicated in each panel is considered (see Table~\ref{tab:scoring} for the 
used radial binnings). Doses are displayed in the arbitrary units adopted in~\cite{verbeek}, \textit{i.e.}, normalized to 
the dose on-axis at \mbox{$z=3$~cm}. At radial distances of \mbox{$r=0$~cm} and \mbox{$r=1.52$~cm} 
(panels~\ref{fig:verbeek_225MeV}a and~\ref{fig:verbeek_225MeV}b), where the dose is high, the remarkable agreement between 
the experimental dose and FLUKA~\mbox{v4-3.4} simulation results is retained in~\mbox{v4-4.0}. At \mbox{$r=2.29$~cm} 
(panel~\ref{fig:verbeek_225MeV}c), where the dose has already dropped by nearly two orders of magnitude, there is a slight 
improvement with FLUKA~\mbox{v4-4.0} at depths beyond $20$~cm, where the dose is nevertheless sizeable. However, substantial 
improvement is obtained at \mbox{$r=7.62$~cm} (panel~\ref{fig:verbeek_225MeV}d) at depths between $25$ and $30$~cm. At these 
depths, better agreement is also obtained at \mbox{$r=9.90$~cm} and \mbox{$r=10.67$~cm} (panels~\ref{fig:verbeek_225MeV}e 
and~\ref{fig:verbeek_225MeV}f), although slight discrepancies remain, which are discussed at the end of 
Section~\ref{sec:move_source}.

\begin{figure*}
\centering
\includegraphics[width=0.95\textwidth]{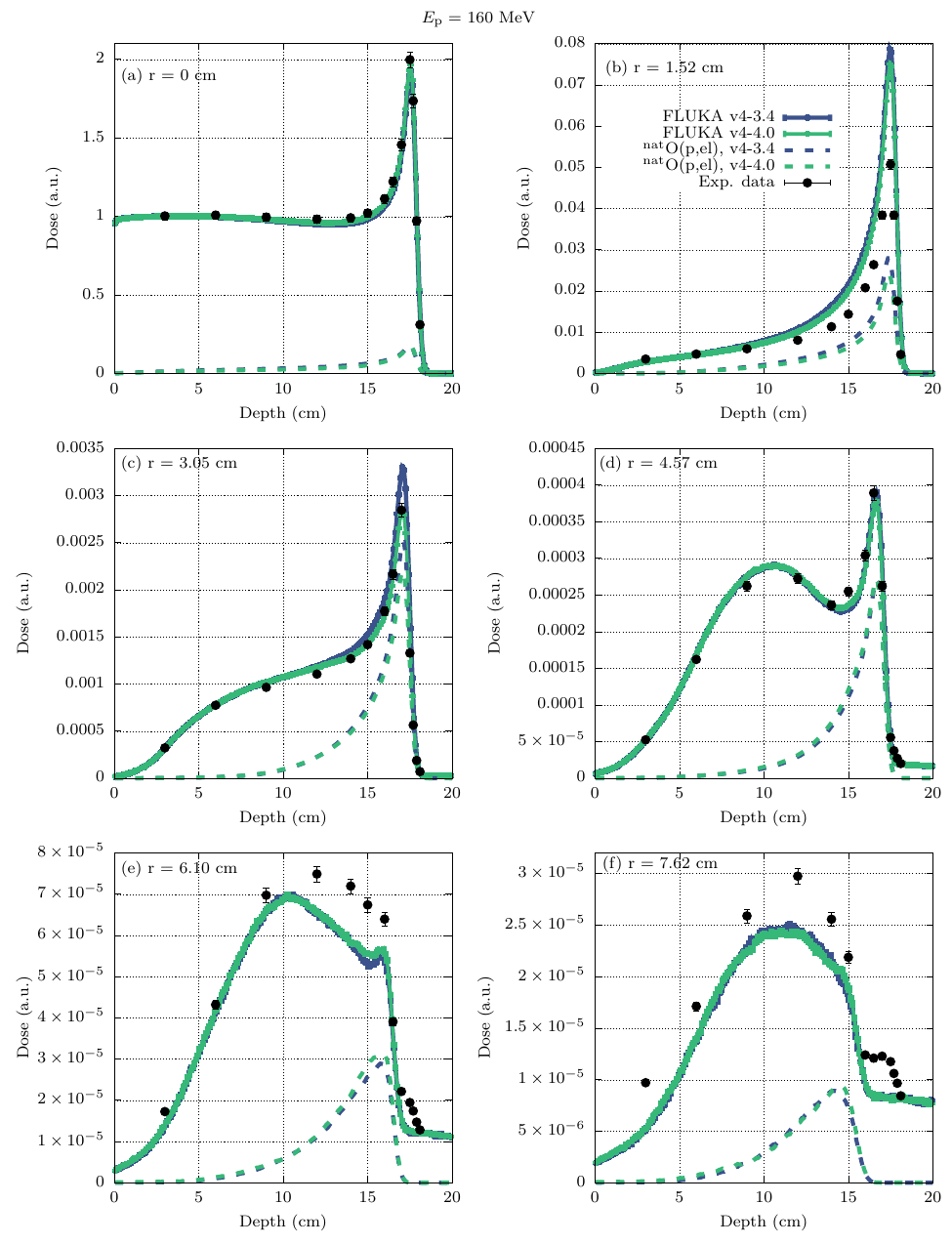}
\caption{Same as Fig.~\ref{fig:verbeek_225MeV} for $160$~MeV protons.}
\label{fig:verbeek_160MeV_noair}
\end{figure*}

The same comparison is displayed in Fig.~\ref{fig:verbeek_160MeV_noair} for the $160$~MeV proton beam case, with its 
corresponding radial binnings (see Table~\ref{tab:scoring}). FLUKA~\mbox{v4-4.0} simulation results at \mbox{$r=0$~cm}, 
\mbox{$r=1.52$~cm}, and \mbox{$r=4.57$~cm} (panels~\ref{fig:verbeek_160MeV_noair}a, \ref{fig:verbeek_160MeV_noair}b, 
and~\ref{fig:verbeek_160MeV_noair}d) are slightly, yet noticeably, closer to the experimental absorbed dose than those 
of~\mbox{v4-3.4}; for \mbox{$r=1.52$~cm} (panel~\ref{fig:verbeek_160MeV_noair}b), the considerable overestimation of the 
experimental absorbed dose in FLUKA~\mbox{v4-3.4} is only minimally reduced in~\mbox{v4-4.0}. A similar disagreement 
with the experimental absorbed doses was found in~\cite{verbeek}, for simulations performed with other MC codes. 
Instead, at \mbox{$r=3.05$~cm} (panel~\ref{fig:verbeek_160MeV_noair}c), the agreement is substantially improved with 
FLUKA~\mbox{v4-4.0}. However, at \mbox{$r=6.10$~cm} and \mbox{$r=7.62$~cm} (panels~\ref{fig:verbeek_160MeV_noair}e 
and~\ref{fig:verbeek_160MeV_noair}f), the differences between the experimental absorbed dose and FLUKA~\mbox{v4-3.4} 
estimates, which are further scrutinized below, are not improved with~\mbox{v4-4.0}: the simulated absorbed doses remain 
much lower than the experimental dose. Moreover, the additional feature at depths between $15$ and $20$~cm is not captured 
at all, especially for \mbox{$r=7.62$~cm} (see Section~\ref{sec:move_source}).

\begin{figure*}
\centering
\includegraphics[width=0.95\textwidth]{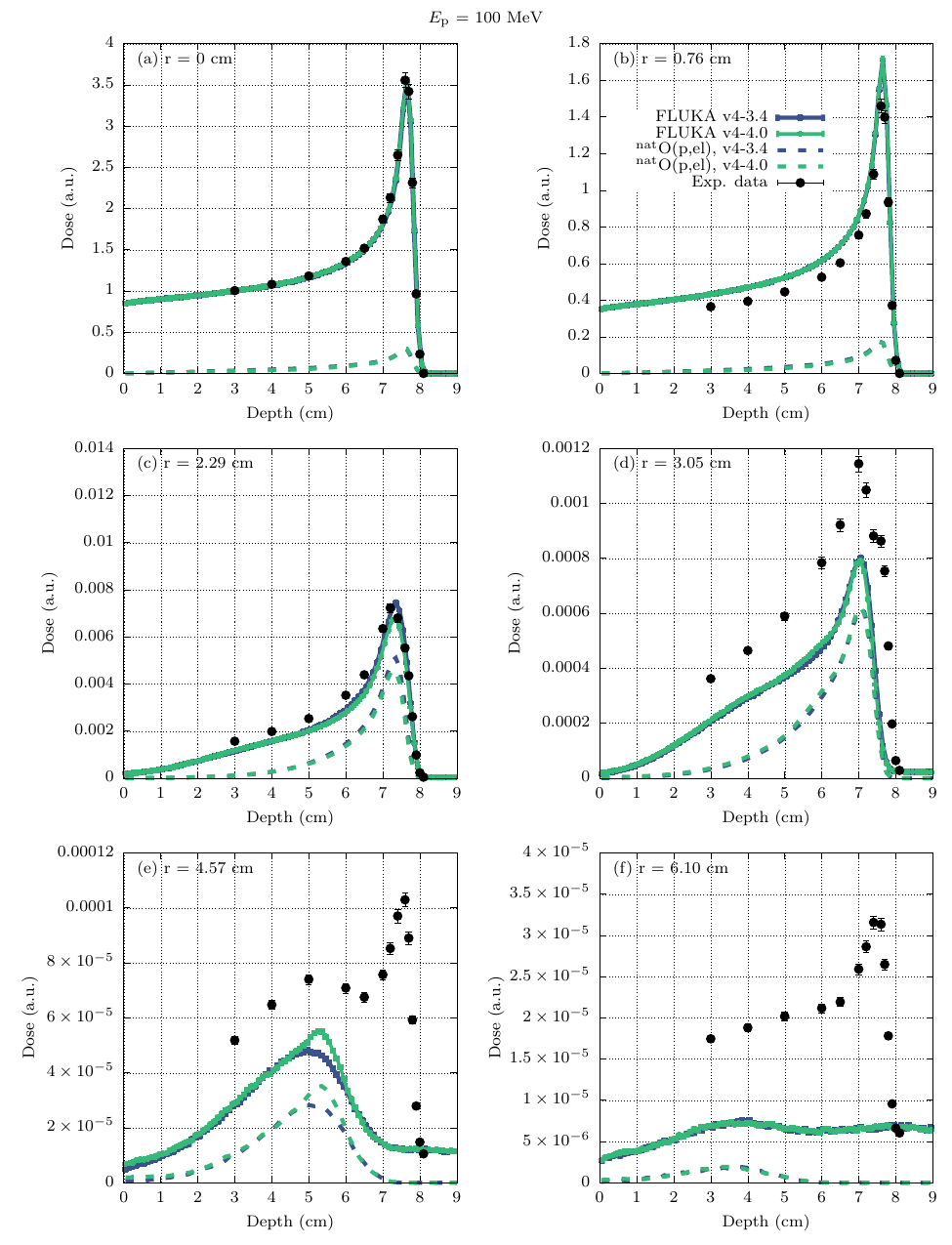}
\caption{Same as Fig.~\ref{fig:verbeek_225MeV} for $100$~MeV protons.}
\label{fig:verbeek_100MeV_noair}
\end{figure*}

Lastly, Fig.~\ref{fig:verbeek_100MeV_noair} depicts the same benchmark for the $100$~MeV proton beam case, with its 
corresponding radial binnings (see Table~\ref{tab:scoring}). The performances of FLUKA~\mbox{v4-3.4} and~\mbox{v4-4.0} 
are similar throughout all radial distances considered here. The agreement with experimental absorbed doses at 
\mbox{$r=$\{$0$, $0.76$, $2.29$\}}~cm (panels~\ref{fig:verbeek_100MeV_noair}a,~\ref{fig:verbeek_100MeV_noair}b, 
and~\ref{fig:verbeek_100MeV_noair}c) is reasonable. Instead, at \mbox{$r=$\{$3.05$, $4.57$, $6.10$\}}~cm 
(panels~\ref{fig:verbeek_100MeV_noair}d,~\ref{fig:verbeek_100MeV_noair}e, and~\ref{fig:verbeek_100MeV_noair}f) 
considerable discrepancies remain.

Before further investigating in Section~\ref{sec:move_source} the origin of the shortcomings at \mbox{$r=6.10$~cm} and 
\mbox{$r=7.62$~cm} for the $160$~MeV case and at \mbox{$r=$\{$3.05$, $4.57$, $6.10$\}}~cm for the $100$~MeV 
case, the physical origin of the improvements observed with FLUKA~\mbox{v4-4.0} is substantiated. For this purpose, 
FLUKA's particle-latching capabilities have been exploited to disentangle the contribution of various kinds of particle 
histories to the $r$-$z$ dose maps. For the three considered energies, 
Figs.~\ref{fig:verbeek_225MeV},~\ref{fig:verbeek_160MeV_noair}, and~\ref{fig:verbeek_100MeV_noair} additionally depict 
the contribution to the total absorbed dose from particle histories where protons underwent a nuclear elastic 
scattering event on oxygen in FLUKA~\mbox{v4-3.4} (dashed blue curve) and in~\mbox{v4-4.0} (dashed green curve), 
respectively. The magnitude of the dashed curves confirms that proton nuclear elastic scattering plays a dominant role 
at large radial distances. Even on axis it is seen to have a relevant contribution towards the end of the proton range. 
The significant increase in the FLUKA~\mbox{v4-4.0} dose, helping to narrow the gap with experimental absorbed doses at 
large radial distances and large depths for the $160$~MeV and $225$~MeV cases, indeed stems from a more accurate 
treatment of proton nuclear elastic scattering in FLUKA~\mbox{v4-4.0}. In particular, nuclear elastic scattering events 
of protons on oxygen are described more accurately, since the FLUKA~\mbox{v4-4.0} model relies on a fit to experimental 
differential cross sections for light nuclei~\cite{exfor,zerkin}. For the $100$~MeV proton beam case, a similar 
behaviour is found.

\begin{table}[h!]
\caption{Relative root mean square deviation, Eq.~\eqref{eq:fof}, of the absorbed dose in water under $100$~MeV, 
$160$~MeV, and $225$~MeV proton irradiation, simulated with FLUKA~\mbox{v4-3.4} and FLUKA~\mbox{v4-4.0}.}
\label{tab:delta}
\centering
\begin{tabular}{|c|c|c|c|}
\hline
\multirow{3}{*}{$E~(\text{MeV})$} & \multirow{3}{*}{$r_j~(\text{cm})$} & \multicolumn{2}{c|}{$\delta(r_j)$} \\ \cline{3-4}
  &  & \makecell{FLUKA \\ \textbf{v4-3.4}} & \makecell{FLUKA \\ \textbf{v4-4.0}} \\ \hline
\multirow{6}{*}{\textbf{100}} 
  & $0$    & $0.0274$  & $0.0258$  \\ 
  & $0.76$ & $0.0805$  & $0.0823$  \\ 
  & $2.29$ & $0.0544$  & $0.0738$  \\ 
  & $3.05$ & $0.3200$  & $0.3299$  \\ 
  & $4.57$ & $0.5639$  & $0.5617$  \\ 
  & $6.10$ & $0.6209$  & $0.6196$  \\ \hline

\multirow{6}{*}{\textbf{160}} 
  & $0$    & $0.0304$  & $0.0269$  \\ 
  & $1.52$ & $0.2938$  & $0.2598$  \\ 
  & $3.05$ & $0.0862$  & $0.0327$  \\ 
  & $4.57$ & $0.0332$  & $0.0293$  \\ 
  & $6.10$ & $0.1231$  & $0.1128$  \\ 
  & $7.62$ & $0.1652$  & $0.1663$  \\ \hline

\multirow{6}{*}{\textbf{225}} 
  & $0$     & $0.0153$ & $0.0119$  \\ 
  & $1.52$  & $0.0360$ & $0.0332$  \\ 
  & $2.29$  & $0.0726$ & $0.0504$  \\ 
  & $7.62$  & $0.0834$ & $0.0296$  \\ 
  & $9.90$  & $0.1098$ & $0.0813$  \\ 
  & $10.67$ & $0.1214$ & $0.1029$  \\ \hline
\end{tabular}
\end{table}

To quantify the agreement between doses simulated with FLUKA~\mbox{v4-3.4}/\mbox{v4-4.0} and the experimental 
absorbed dose at each radial distance, the relative root mean square deviation is adopted:
\begin{equation}
\label{eq:fof}
 \delta(r_j) = \cfrac{ \sqrt{ \cfrac{1}{N} \sum \limits_{i=1}^{N} \left( D_{ij}^{\mathrm{FLUKA}} - D_{ij}^{\mathrm{exp}} \right)^2}}
                     { D_{j}^{\mathrm{exp,max}} - D_{j}^{\mathrm{exp,min}} }
               ,
\end{equation}
where $N$ is the number of experimental points along the $z$ axis at each radial distance $r_j$, 
\mbox{$j=\{1,2,3,4,5,6\}$}, while $D_{ij}^{\mathrm{FLUKA}}$ and $D_{ij}^{\mathrm{exp}}$ are the FLUKA-simulated and 
the experimental absorbed dose at each $z_i$ and $r_j$, respectively. The maximum and minimum experimental absorbed 
dose at each $r_j$ are denoted by $D_{j}^{\mathrm{exp,max}}$ and $D_{j}^{\mathrm{exp,min}}$, respectively. 
Table~\ref{tab:delta} displays $\delta(r_j)$ evaluated for both FLUKA~\mbox{v4-3.4} and~\mbox{v4-4.0}. For the 
$225$~MeV proton beam case, $\delta(r_j)$ confirms that better agreement with the experimental absorbed dose is 
achieved at all radial distances with FLUKA~\mbox{v4-4.0} than with~\mbox{v4-3.4}. Likewise for the $160$~MeV proton 
beam case, except for \mbox{$r=7.62$~cm}, where $\delta(r_j)$ are comparable for both versions of the code. For the 
$100$~MeV case, both FLUKA versions have similar $\delta(r_j)$.

\section{Source term for the 100~MeV and \mbox{160~MeV} proton beams}
\label{sec:move_source}

As shown in Section~\ref{sec:simulations_noair}, the Fermi-Eyges description of the proton beam at the entrance of the water 
phantom is sufficiently effective for the purposes of the MC simulation of the experimental $r$-$z$ dose maps~\cite{verbeek} 
for protons of $225$~MeV, at all radial distances considered here, as well as for protons of $160$~MeV and $100$~MeV, but 
only on axis. Instead, at large radial distances, several underestimations and missing features are witnessed for the latter 
two proton energies. Similar difficulties were reported in~\cite{verbeek}. In this section, the underlying issues are 
substantiated and an effective prescription to overcome them is proposed.

\begin{figure*}
\centering
\includegraphics[width=0.7\textwidth]{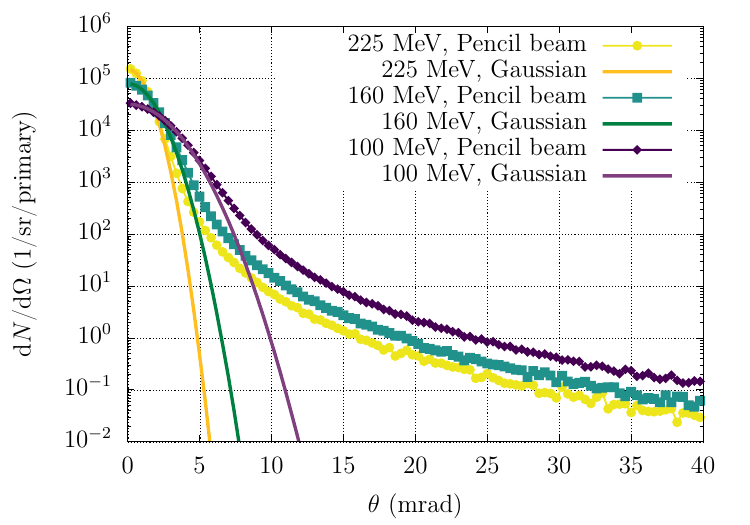}
\caption{Comparison between FLUKA sampled angular distributions of pencil proton beams after $50$~cm of air vs.\ fitted Gaussian
distributions for proton energies of $225$~MeV (yellow circles), $160$~MeV (green squares), and $100$~MeV (purple diamonds).}
\label{fig:gaussian}
\end{figure*}

The Fermi-Eyges theory outlined in Section~\ref{sec:source} is effective at capturing the dominating Gaussian core of 
the angular distribution of protons impinging on the water phantom, stemming from multiple Coulomb scattering (MCS) as 
the incoming protons traverse air on their path from the beamline to the water phantom. However, it does not account 
for large-scattering-angle contributions (be it from Coulomb or nuclear elastic scattering) extending to angles beyond 
those of the Gaussian core. To elucidate this point in a simplified way, Fig.~\ref{fig:gaussian} displays angular 
distributions of $225$~MeV (yellow circles), $160$~MeV (green squares), and $100$~MeV (purple diamonds) proton pencil 
beams after traversing $50$~cm of air, sampled with FLUKA. These curves indeed exhibit a dominating Gaussian structure 
around the incoming direction, but also the expected larger-scattering-angle tail. The solid curves of matching colours 
display the extent to which a Gaussian fit manages to reproduce these angular distributions. This analysis suggests 
that the Fermi-Eyges source model is not sufficiently accurate to capture large-$r$ features for the $160$~MeV and 
$100$~MeV cases, hence the difficulties reported in Figs.~\ref{fig:verbeek_160MeV_noair} 
and~\ref{fig:verbeek_100MeV_noair}.

\begin{figure*}
\centering
\includegraphics[width=\textwidth]{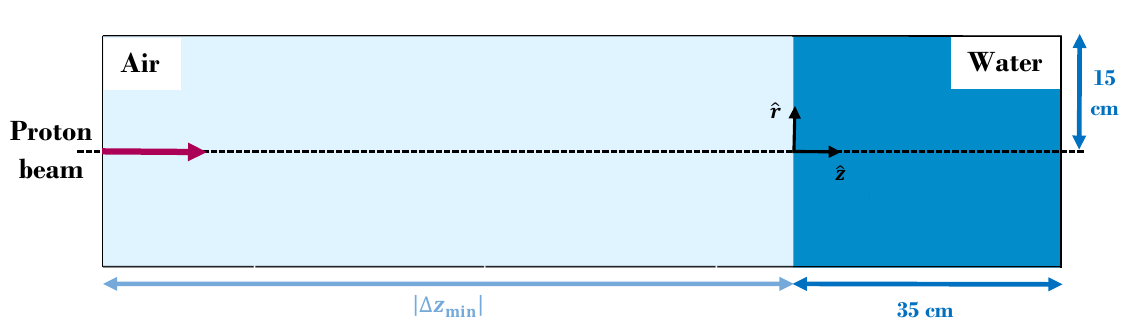}
\caption{Schematic FLUKA simulation setup for the $100$~MeV and $160$~MeV proton beams, consisting of the water phantom and
the impinging proton beam after traversing an air layer.}
\label{fig:geom}
\end{figure*}

\begin{figure*}
\centering
\includegraphics[width=0.9\textwidth]{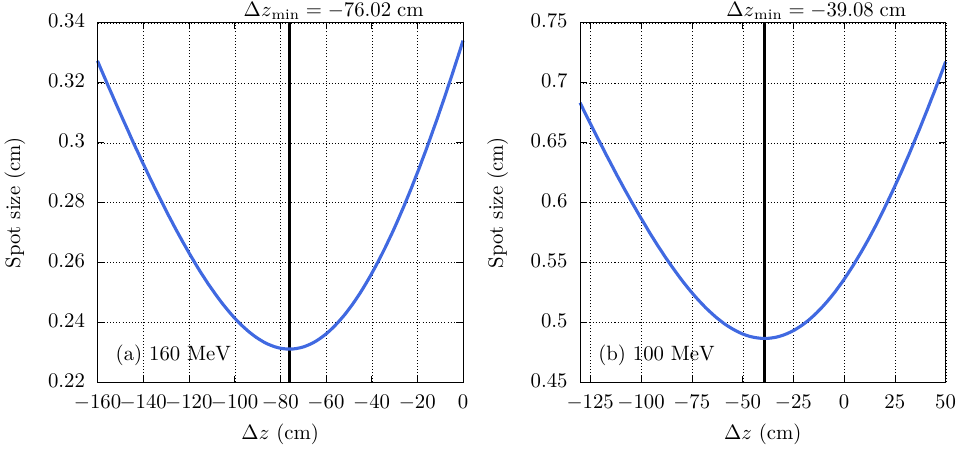}
\caption{Beam spot size as a function of source displacement.}
\label{fig:spot_size}
\end{figure*}

To explicitly simulate deflections beyond the Fermi-Eyges Gaussian core, an air layer has been added in front of the 
water phantom, as depicted in Fig.~\ref{fig:geom}, in which protons may undergo nuclear elastic scattering. The 
proton source must therefore be retracted and the corresponding Fermi-Eyges beam parameters must be determined. Since 
these are known at the entrance of the water phantom~\cite{verbeek}, the new beam parameters can be determined 
following~\cite{GottschalkFE}, assuming that air is a weak scatterer, as:
\begin{equation}
\begin{aligned}
\label{eq:a_parameters}
   A_0(z) & = A_0(z_1) + T \Delta z\\
   A_1(z) & = A_1(z_1) + A_0(z_1) \Delta z + T \frac{\Delta z^2}{2} \\
   A_2(z) & = A_2(z_1) + 2A_1(z_1) \Delta z + \\
            & + A_0(z_1) \Delta z^2 + T \frac{\Delta z^3}{3},
\end{aligned}
\end{equation}
where $\Delta z = z - z_1$; $z_1$ is the initial position of the source at the entrance of the water 
phantom, \textit{i.e.}, $0$~cm, and $z<0$ is the new position of the source. The beam parameters at the entrance of the 
water phantom~\cite{verbeek}, \textit{i.e.}, $A_0(z_1)$, $A_1(z_1)$, and $A_2(z_1)$, are summarized in 
Table~\ref{tab:simulation_param}. The scattering power~\cite{GottschalkFE}
\begin{equation}
\label{eq:scatt_power}
   T = \frac{\mathrm{d} \langle \theta^2 \rangle}{\mathrm{d}z},
\end{equation}
assumed constant for air, has been estimated as follows. The angular distributions of $100$~MeV and
$160$~MeV protons traversing air slabs of various thicknesses have been scored with FLUKA. The mean squared projected MCS
angle has been evaluated for each considered thickness and $T$ has been estimated using a finite difference
approximation of Eq.~\eqref{eq:scatt_power}.
The beam spot size is:
\begin{equation}
\label{eq:lat_displacement_var}
   \sigma_x(z) = \sqrt{2A_2(z)} .
\end{equation}
Figure~\ref{fig:spot_size} displays the beam spot size as a function of the source displacement $\Delta z$ for the $160$~MeV 
(a) and $100$~MeV (b) proton beam cases, respectively. The beam spot sizes attain a minimum at \mbox{${\Delta 
z}_\mathrm{min} = z_\mathrm{min}-z_1= -76.02$~cm} for $160$~MeV and $-39.08$~cm for $100$~MeV, indicated by the black 
vertical lines in Figs.~\ref{fig:spot_size}a and~\ref{fig:spot_size}b. The parameters of the retracted proton beams at the 
position of the minima are listed in Table~\ref{tab:160MeV_param_min}. Beyond these minima, the Fermi-Eyges theory cannot be 
used to move the source farther back since this prescription only accounts for the beam spread~\cite{Gottschalk}, and not 
for beam focusing effects. Nevertheless, one can account for air layers thicker than $|\Delta z_\mathrm{min}|$ by scaling 
the density of an air layer of thickness $|\Delta z_\mathrm{min}|$ as
\begin{equation}
    \rho_\mathrm{equiv} = \rho_\mathrm{air} \frac{|{\Delta z}|}{|{\Delta z}_\mathrm{min}|},
    \label{eq:rhoequiv}
\end{equation}
where $\rho_\mathrm{air}=1.20479 \times 10^{-3} $ g/$\mathrm{cm}^3$~\cite{Zhang2010,Burin2023} is the density of dry 
air at sea level and $|\Delta z|>|\Delta z_\mathrm{min}|$ is the thickness for which best agreement between simulated 
and experimental $r$-$z$ dose maps is obtained (found by iteration). In this way, an air layer thicker than $|\Delta 
z_\mathrm{min}|$ can be effectively simulated without retracting the spatial position of the source beyond the domain 
of applicability of the Fermi-Eyges prescription.
Due to the source displacement, the simulated proton beam loses energy in the additional air layer. Thus, the position of
the Bragg peak changes. To obtain it at the correct depth, the beam energy is effectively adjusted as~\cite{Bortfeld}:
\begin{equation}
\label{eq:energy_correction}
    E^{\prime} = \left( E^p - \frac{\Delta z}{\alpha} \right)^{1/p},
\end{equation}
where $E$ is the beam energy at the water phantom entrance, while $p=1.7589$ and $\alpha=2.194$~MeV$^{-p}$~cm are parameters 
obtained by fitting the energy-range power law $R=\alpha E^p$, where $R$ is the range of protons in air~\cite{nist,PSTAR}. 
Additionally, since the straggling in air is negligible, the energy spread $\sigma_E$ of the proton beam at the entrance of 
the water phantom, summarized in Table~\ref{tab:simulation_param}, remains unaltered. The adjusted proton beam energy 
$E^{\prime}$, the equivalent air density $\rho_\mathrm{equiv}$, and the thickness $|{\Delta z}|$ of the air layer needed to 
obtain $\rho_\mathrm{equiv}$ are outlined in Table~\ref{tab:160_new_param_mod}.

\begin{table}
\caption{Parameters of the retracted source at the position at which the beam spot size minimum 
is attained for the $100$~MeV and $160$~MeV proton beams.}
\label{tab:160MeV_param_min}
\begin{center}
\begin{tabular}{|l|c|c|}
\hline & \makecell{\textbf{100} \\ \textbf{MeV}} & \makecell{\textbf{160} \\ \textbf{MeV}} \\
\hline $T$~(rad$^2$/cm)                     & $12 \times 10^{-8}$ & $4.8 \times 10^{-8}$ \\
\hline ${\Delta z}_\mathrm{min}$~(cm)       & $-39.08$            & $-76.02$ \\
\hline $\sqrt{2A_2(z_\mathrm{min})}$ (cm)   & $0.484$             & $0.216$ \\
\hline $2A_1(z_\mathrm{min})$ (cm mrad)     & $0.0916$            & $0.133$ \\
\hline $\sqrt{2A_0(z_\mathrm{min})}$ (mrad) & $5.17$              & $2.32$  \\
\hline
\end{tabular}
\end{center}
\end{table}

\begin{table}
\caption{Adjusted proton beam energy $E^{\prime}$ , thickness $|\Delta z|$ of the air layer, and 
equivalent density as per Eq.~\eqref{eq:rhoequiv} employed for the $100$~MeV and $160$~MeV cases.}
\label{tab:160_new_param_mod}
\begin{center}
\begin{tabular}{|l|c|c|}
\hline & \makecell{\textbf{100} \\ \textbf{MeV}} & \makecell{\textbf{160} \\ \textbf{MeV}} \\
\hline $E^{\prime}$ (MeV)                        & $101.167$                & $161.450$ \\
\hline $|{\Delta z}|$ (cm)                       & $130$                    & $220$ \\
\hline $\rho_\mathrm{equiv}$ (g/$\mathrm{cm}^3$) & $4.00755 \times 10^{-3}$ & $3.48663 \times 10^{-3}$ \\
\hline
\end{tabular}
\end{center}
\end{table}

\begin{figure*}
\centering
\includegraphics[width=0.95\textwidth]{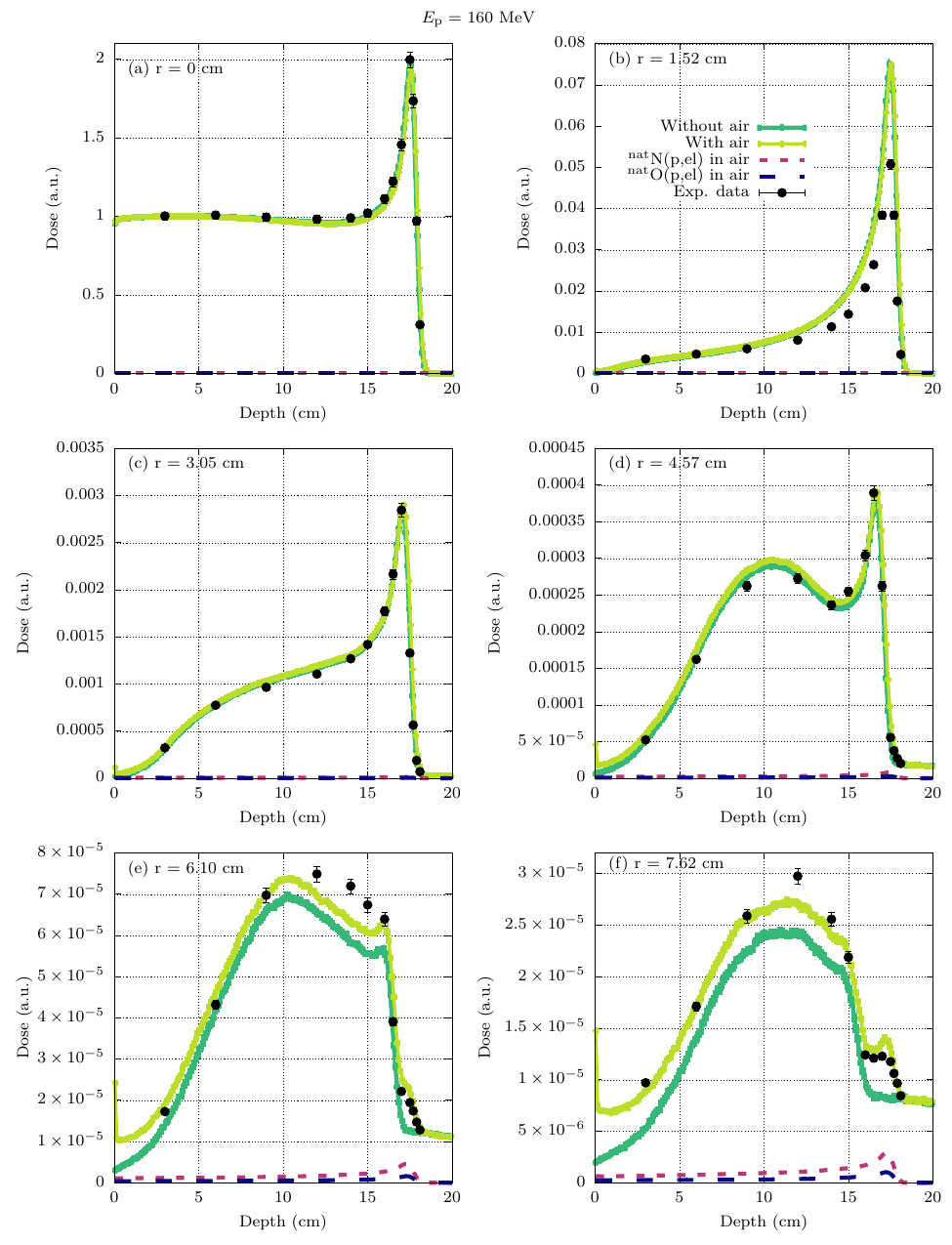}
\caption{Absorbed dose in arbitrary units as a function of depth by $160$~MeV protons in water scored with 
FLUKA~\mbox{v4-4.0} without (green) and with (yellow) the air layer. The black dots represent experimental absorbed 
doses~\cite{verbeek}. The dashed curves correspond to contributions from particle histories where protons underwent a 
nuclear elastic scattering in air on nitrogen (magenta) and on oxygen (blue). In panels (a)-(c) these
contributions nearly vanish at the scale of the figure.}
\label{fig:verbeek_160MeV_air_effect}
\end{figure*}

\begin{figure*}
\centering
\includegraphics[width=0.95\textwidth]{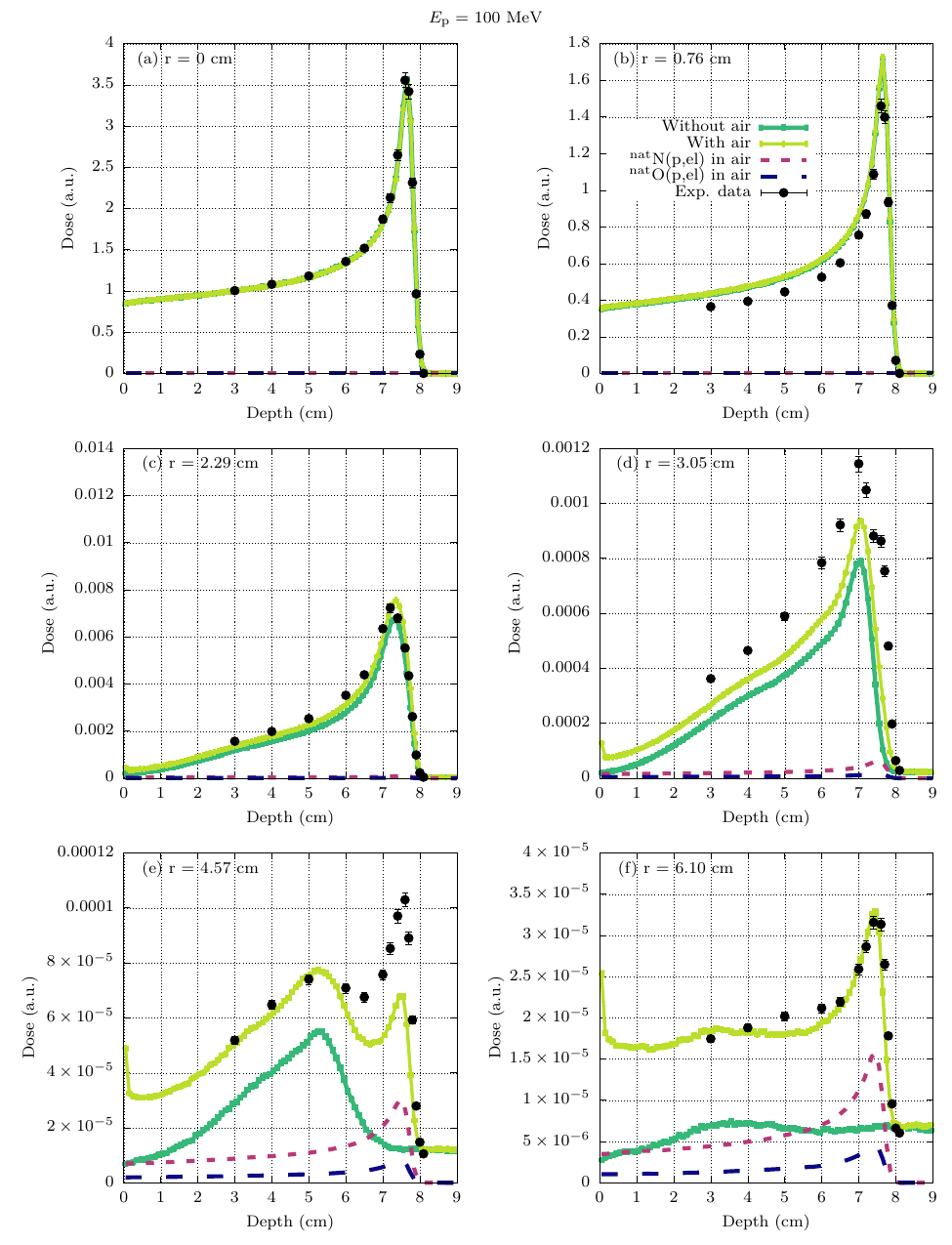}
\caption{Same as Fig.~\ref{fig:verbeek_160MeV_air_effect} for $100$~MeV.}
\label{fig:verbeek_100MeV_air_effect}
\end{figure*}

Figures~\ref{fig:verbeek_160MeV_air_effect} and~\ref{fig:verbeek_100MeV_air_effect} show the drastic effect of the 
additional air layer in front of the water phantom for $160$~MeV and $100$~MeV, respectively, where it matters most, 
\textit{i.e}, at large radial distances from the beam axis. Experimental absorbed doses~\cite{verbeek} are displayed in 
black symbols, while the FLUKA~\mbox{v4-4.0} estimates with and without the additional air layer are shown in yellow and 
green curves, respectively. The dashed magenta and blue curves correspond to particle histories where protons underwent a 
nuclear elastic scattering in air, on nitrogen and on oxygen, respectively. These events occuring in the additional air 
layer prove to be essential: without their contribution (alongside the contribution from nuclear reactions of protons in 
air) the simulated dose is overall much lower than the experimental one. In particular, for $160$~MeV the features exhibited 
by the experimental absorbed dose at \mbox{$r=6.10$~cm} (panel~\ref{fig:verbeek_160MeV_air_effect}e) and \mbox{$r=7.62$~cm} 
(panel~\ref{fig:verbeek_160MeV_air_effect}f) at depths around $15$--$20$~cm are not captured at all by the simulated dose 
when the air layer is missing (green curves). Instead, with the latter properly accounted for, these additional features 
(originating from the nuclear elastic scattering of protons on nitrogen and on oxygen in air) are correctly reproduced 
(yellow curves). The agreement at the first four radial distances 
(panels~\ref{fig:verbeek_160MeV_air_effect}a,~\ref{fig:verbeek_160MeV_air_effect}b, 
~\ref{fig:verbeek_160MeV_air_effect}c,~\ref{fig:verbeek_160MeV_air_effect}d) has been generally preserved while retracting 
the source. The same holds for $100$~MeV on axis (panel~\ref{fig:verbeek_100MeV_air_effect}a) and at \mbox{$r=0.76$~cm} 
(panel~\ref{fig:verbeek_100MeV_air_effect}b). At \mbox{$r=2.29$~cm} (panel~\ref{fig:verbeek_100MeV_air_effect}c), there is a 
mild improvement which helps to close the gap between the FLUKA-simulated and the experimental absorbed dose. However, at 
\mbox{$r=3.05$~cm} (panel~\ref{fig:verbeek_100MeV_air_effect}d) and \mbox{$r=4.57$~cm} 
(panel~\ref{fig:verbeek_100MeV_air_effect}e) this is not the case: the increase in the FLUKA-simulated dose is insufficient 
to match the experimental absorbed dose (see below). The most remarkable enhancement, leading to a satisfactory agreement 
with the experimental absorbed dose, is seen at \mbox{$r=6.10$~cm} (panel~\ref{fig:verbeek_100MeV_air_effect}f). At this 
radial distance, proton nuclear elastic scattering on nitrogen in the air layer in front of the water phantom proves to have 
a massive contribution, as illustrated by the dashed magenta curve in panel~\ref{fig:verbeek_100MeV_air_effect}f.

The mean free paths for the nuclear elastic scattering of $100$~MeV, $160$~MeV, and $225$~MeV protons in air are $2.37 
\times 10^4$~cm, $6.75 \times 10^4$~cm, and $2.34 \times 10^5$~cm, respectively. While for protons of $100$~MeV and 
$160$~MeV the mean free paths are comparable, for $225$~MeV protons it is one order of magnitude larger. This explains why 
the Fermi-Eyges source characterization was effective for the $225$~MeV proton beam and not for $100$~MeV and $160$~MeV 
proton beams.

The features at $z=0$ observed with the inclusion of the air layer in Figs.~\ref{fig:verbeek_160MeV_air_effect} 
and~\ref{fig:verbeek_100MeV_air_effect}, notably at large $r$, are due to $\delta$ rays generated by secondary particles in 
the air region preceding the water phantom. These are travelling from a low-density medium (air) to a higher-density medium 
(water), where they have a much shorter range. Thus, the $\delta$ rays generated in air deposit their energy over a shorter 
distance in water and lead to the observed localized dose deposition at the entrance of the phantom.

\begin{table}[h!]
\caption{Relative root mean square deviation, Eq.~\eqref{eq:fof}, of the absorbed dose in water under $100$~MeV
and $160$~MeV proton irradiation, simulated with FLUKA~\mbox{v4-4.0} with and without the additional air layer.}
\label{tab:delta_air}
\centering
\begin{tabular}{|c|c|c|c|}
\hline
\multirow{2}{*}{$E~(\text{MeV})$} & \multirow{2}{*}{$r_j~(\text{cm})$} & \multicolumn{2}{c|}{$\delta(r_j)$} \\ \cline{3-4}
  &  & \textbf{Without air} & \textbf{With air} \\ \hline

\multirow{6}{*}{\textbf{100}}
  & $0$    & $0.0258$ & $0.0294$ \\
  & $0.76$ & $0.0823$ & $0.0890$ \\
  & $2.29$ & $0.0738$ & $0.0460$ \\
  & $3.05$ & $0.3299$ & $0.2301$ \\
  & $4.57$ & $0.5617$ & $0.2357$ \\
  & $6.10$ & $0.6196$ & $0.1163$ \\ \hline

\multirow{6}{*}{\textbf{160}}
  & $0$    & $0.0269$ & $0.0616$ \\
  & $1.52$ & $0.2598$ & $0.2884$ \\
  & $3.05$ & $0.0327$ & $0.0718$ \\
  & $4.57$ & $0.0293$ & $0.0524$ \\
  & $6.10$ & $0.1128$ & $0.0625$ \\
  & $7.62$ & $0.1663$ & $0.0523$ \\ \hline

\end{tabular}
\end{table}

The adopted measure $\delta(r_j)$, Eq.~\eqref{eq:fof}, has been evaluated for the FLUKA~\mbox{v4-4.0} absorbed doses, 
taking into account the additional air layer in front of the water phantom and compared in Table~\ref{tab:delta_air} 
with the FLUKA~\mbox{v4-4.0} estimates from Table~\ref{tab:delta}, where no air layer has been considered. While for 
the $160$~MeV proton beam case, the agreement with the experimental absorbed dose at the first four radial distances 
is to a mild extent deteriorated by the addition of the air layer in front of the water phantom, at \mbox{$r=6.10$~cm} 
and \mbox{$r=7.62$~cm} there is substantial improvement due to the nuclear elastic and inelastic scattering of protons 
in air. Not only is the intensity in the simulated dose higher throughout the depth of the water phantom, but also the 
prominent features at depths beyond $15$~cm are well reproduced. For the $100$~MeV case, at \mbox{$r=0$~cm} and 
\mbox{$r=0.76$~cm} comparable $\delta(r_j)$ are obtained when considering or not the air layer since its effect on 
axis is negligible at this lower proton energy. Instead, at larger radial distances, better agreement with respect to 
the experimental absorbed dose is achieved. The most striking effect of the additional air layer in front of the water 
phantom is observed at \mbox{$r=6.10$~cm}, stemming mainly from the nuclear elastic scattering of protons on nitrogen. 
At this radial distance, the simulated dose increases substantially, matching the experimental absorbed dose.

Residual discrepancies between FLUKA~\mbox{v4-4.0}-simulated and experimental absorbed doses remain for the last two 
radial distances of the $225$~MeV case, as well as for the $160$~MeV case (regardless of the additional air layer). 
Furthermore, for the $160$~MeV case at \mbox{$r=1.52$~cm} the considerable overestimation of the experimental absorbed 
dose by the simulated dose is obtained with both FLUKA versions, and it persists when the air layer is added in front 
of the water phantom. For the $100$~MeV case, differences remain at \mbox{$r=3.05$~cm} and \mbox{$r=4.57$~cm}. Similar 
discrepancies as the aformentioned ones are reported also in the original study~\cite{verbeek} for simulations 
performed with other MC codes. To investigate the extent to which nuclear interactions models could impact these 
discrepancies, the integrated cross section for both nuclear elastic and inelastic scattering of protons on oxygen in 
the energy range of interest ($100$--$225$~MeV) has been varied by a substantial $\pm$$10$\%. This variation has been 
insufficient to close the gap between simulated and experimental absorbed doses (and, moreover, it compromised the 
good agreement at shallow depths). Therefore, the remaining differences may be due to uncertainties in the 
experimental absorbed doses, in the original Fermi-Eyges proton beam parameters (which propagate into uncertainties in 
source parameters in its retracted position at the entrance of the air layer), or due to simplifications of the actual 
experimental beamline layout for simulation purposes.

\section{Conclusions}
\label{sec:conclusions}

The recent inclusion of a new model for the nuclear elastic scattering of protons below $250$~MeV in 
FLUKA~\mbox{v4-4.0}~\cite{serban} implied a necessary reassessment of the code performance for proton dosimetry 
applications, for which this interaction mechanism plays a significant role. The absorbed $r$-$z$ dose distributions 
reported in~\cite{verbeek} constitute a valuable dataset against which to benchmark the performances of this new model. 
Thus, a detailed analysis of these $r$-$z$ dose maps from $225$~MeV, $160$~MeV, and $100$~MeV protons has been carried out 
both with FLUKA~\mbox{v4-3.4}, relying on a legacy model for proton nuclear elastic scattering~\cite{ranft}, and 
with~\mbox{v4-4.0}, including the new model for this interaction mechanism.

For the $225$~MeV proton beam case, this benchmark shows that, while on axis the excellent performances of 
FLUKA~\mbox{v4-3.4} are preserved, an enhanced agreement with the experimental absorbed dose is obtained out-of-field 
with~\mbox{v4-4.0}. An explicit filtering of particle histories contributing to the total simulated dose confirms that these 
notable improvements result from a more accurate description of proton nuclear elastic scattering on oxygen, since the new 
model~\cite{serban} relies on a fit to experimental differential cross sections~\cite{exfor,zerkin} for protons on light 
nuclei.

For the $160$~MeV and $100$~MeV proton beam cases, the new proton nuclear elastic scattering model leads to improvements 
which are, however, insufficient to reproduce the experimental absorbed dose, especially at large radial distances. For 
these two energies, the remaining notable discrepancies are instead due to an incomplete characterization of the proton 
source based on the Fermi-Eyges theory. The latter assumes a Gaussian approximation of the spatial and angular beam 
profiles, disregarding the large-angle contribution stemming from nuclear elastic scattering, which has a more significant 
impact the lower the proton energy. These shortcomings can be overcome by retracting the source and interspersing a layer of 
air in front of the water phantom to explicitly account for large-angle deflections. It has been shown in this work that a 
proper account of nuclear elastic scattering of protons on nitrogen and oxygen in the air layer is indispensable to capture 
the large-$r$ features exhibited by the $r$-$z$ dose maps of $160$~MeV and $100$~MeV protons.

FLUKA's excellent description of dose deposition has not only been preserved on axis, but substantial improvements have been 
achieved out-of-field thanks to the new model for proton nuclear elastic scattering implemented in 
FLUKA~\mbox{v4-4.0}~\cite{serban}. Finally, the importance of an accurate definition of the radiation source term, in 
addition to the need of robust physics models, for detailed Monte Carlo simulation purposes has been highlighted.

\section*{Acknowledgements} 

The authors would like to express their gratitude to L. Brualla for generously providing the experimental radial-depth 
dose maps essential for this work. J.A. de la Torre González, M. Anguiano, and A.M. Lallena acknowledge that this work has 
been partially supported by the Spanish Ministerio de Ciencia y Competitividad (PID2019-104888GB-I00, 
PID2022-137543NB-I00) and the European Regional Development Fund (ERDF).

\section*{Appendix}
\label{app:2D_maps}

\begin{figure*}
\centering
\includegraphics[width=0.95\textwidth]{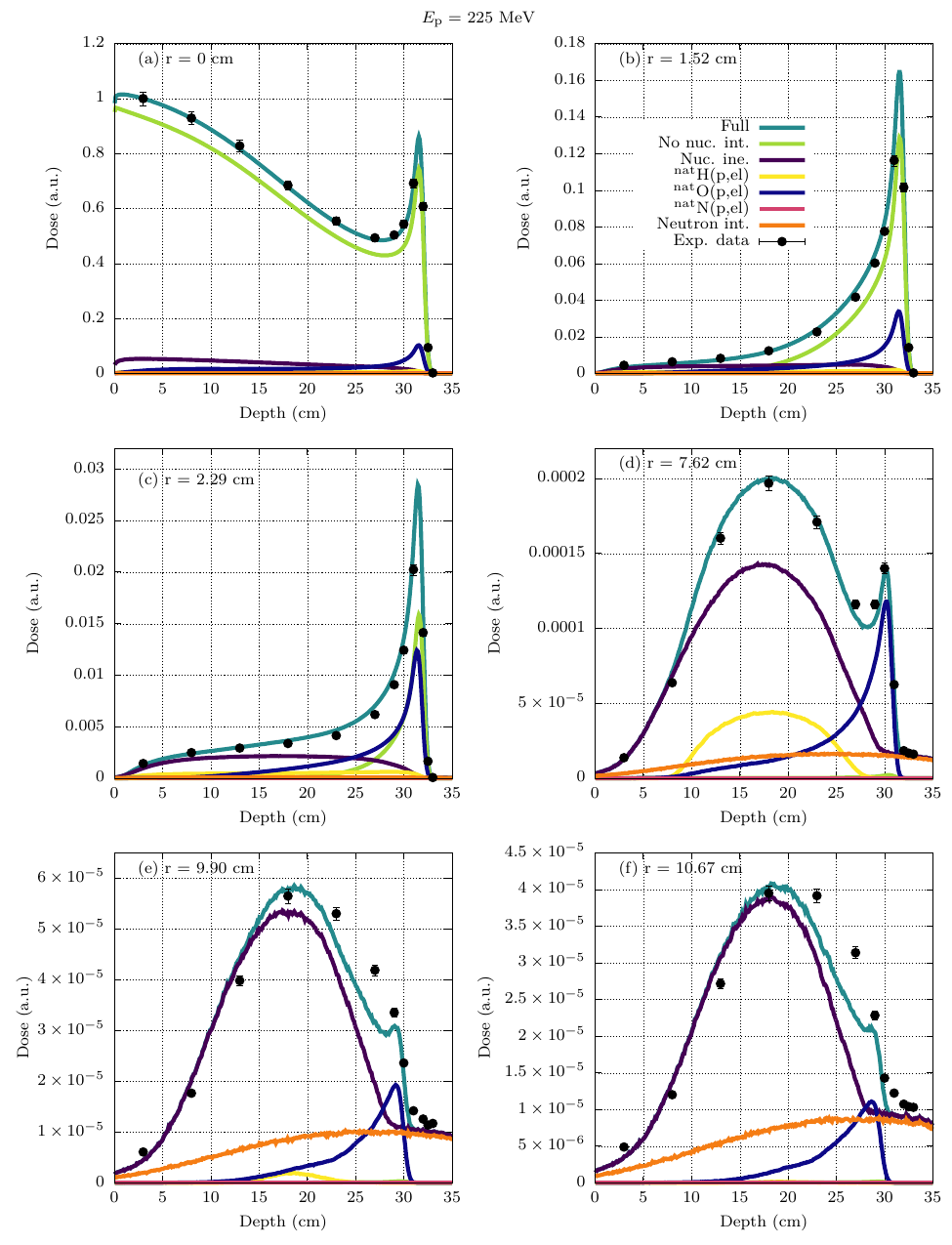}
\caption{FLUKA~\mbox{v4-4.0} absorbed dose in arbitrary units as a function of depth by $225$~MeV protons in water, 
resolved into the contribution of various kind of particle histories (see key and text).}
\label{fig:split_verbeek_225MeV}
\end{figure*}

\begin{figure*}
\centering
\includegraphics[width=0.95\textwidth]{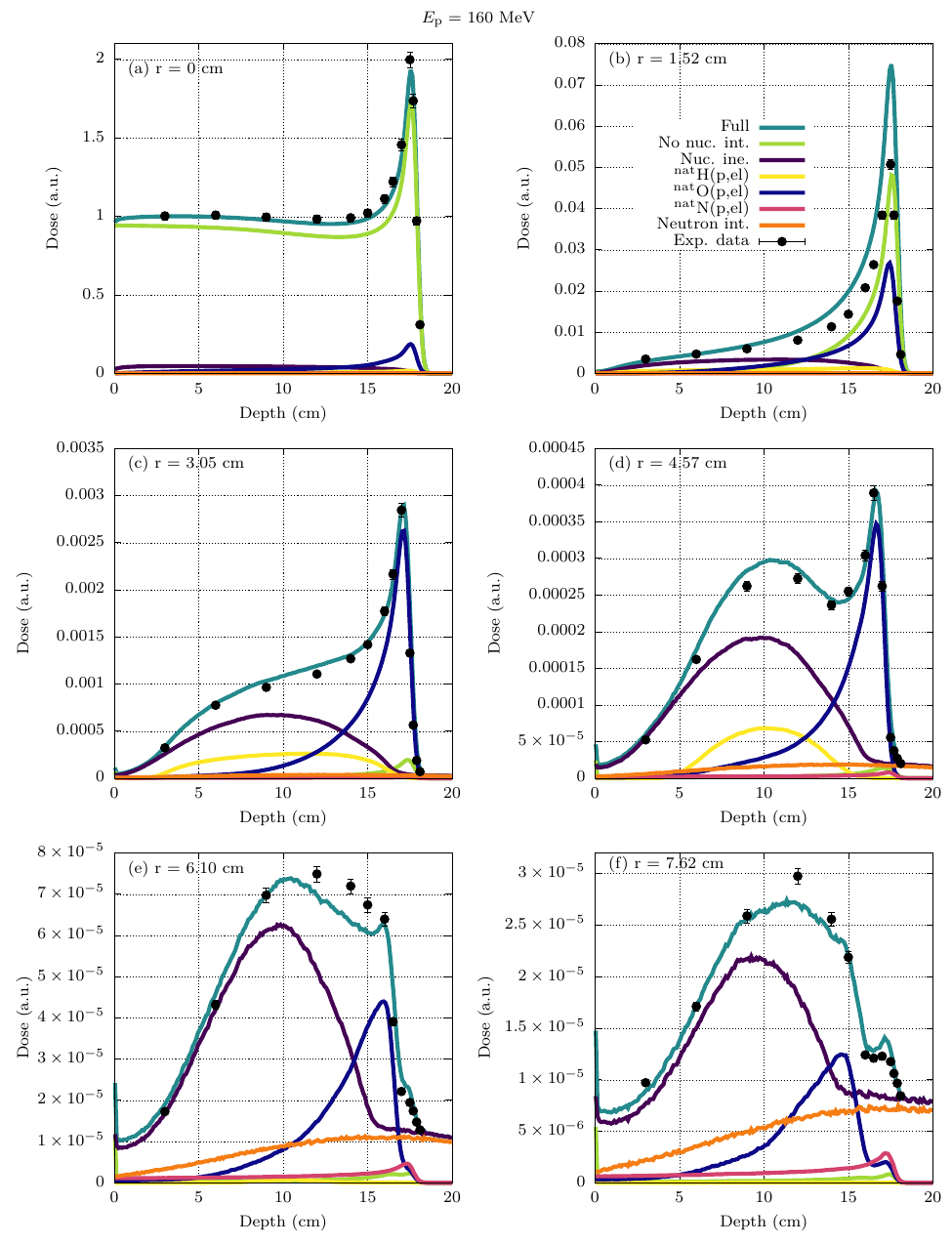}
\caption{Same as Fig.~\ref{fig:split_verbeek_225MeV} for $160$~MeV protons.}
\label{fig:split_verbeek_160MeV}
\end{figure*}

\begin{figure*}
\centering
\includegraphics[width=0.95\textwidth]{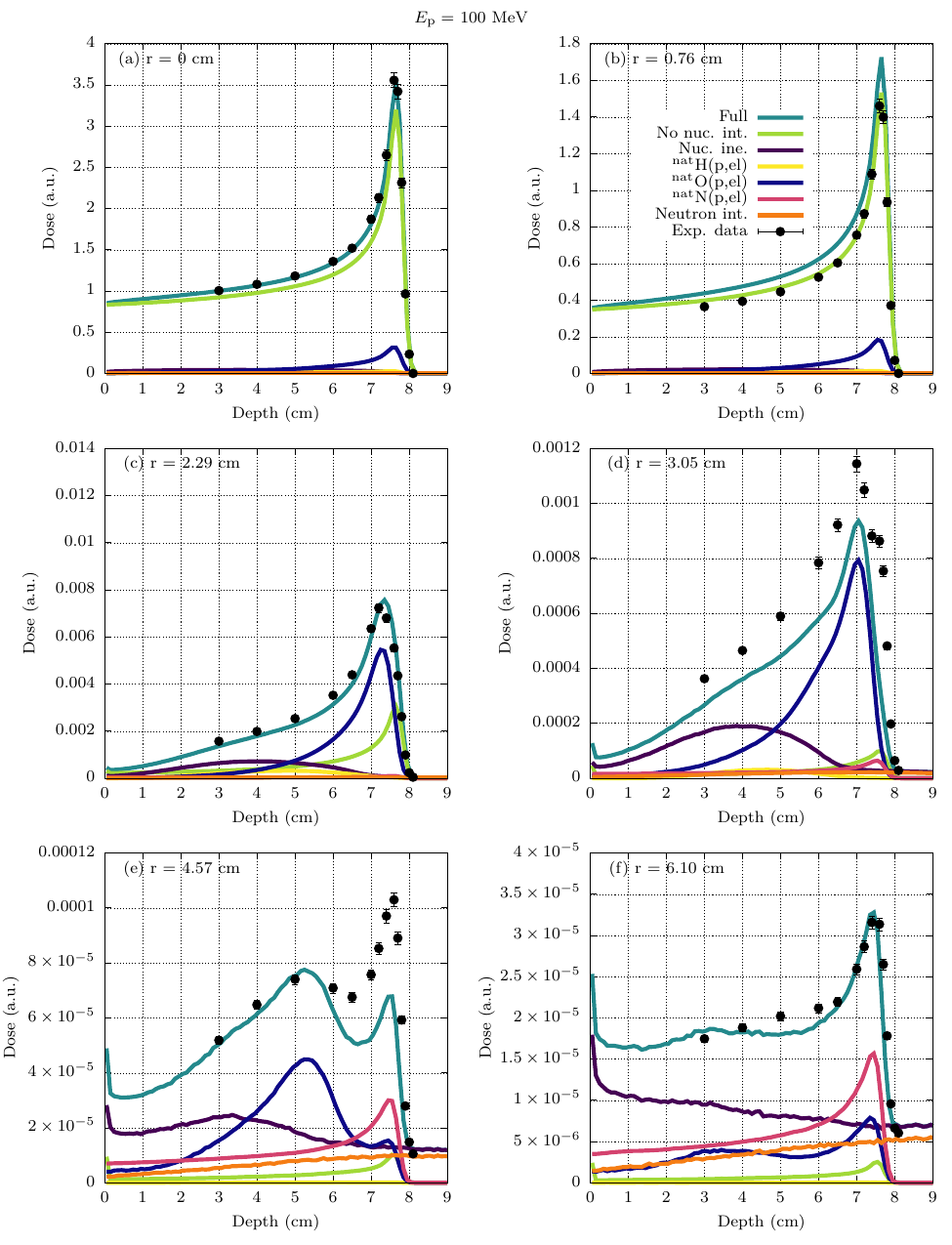}
\caption{Same as Fig.~\ref{fig:split_verbeek_225MeV} for $100$~MeV protons.}
\label{fig:split_verbeek_100MeV}
\end{figure*}

To illustrate how various interaction mechanisms contribute to dose absorption in the setup considered in this work, the 
particle-latching capabilities of FLUKA have been employed to explicitly score the contribution of different kinds of 
particle histories to various regions of the considered $r$-$z$ dose maps. 
Figures~\ref{fig:split_verbeek_225MeV},~\ref{fig:split_verbeek_160MeV}, and~\ref{fig:split_verbeek_100MeV} respectively 
depict for $225$~MeV, $160$~MeV, and $100$~MeV (the latter two with an additional air layer in front of the water phantom, 
as described in Section~\ref{sec:move_source}) the total dose scored with FLUKA~\mbox{v4-4.0} (teal). The total dose has 
been further resolved into the contributions from particle histories where protons have undergone a nuclear reaction 
(purple), from which the contribution resulting from neutron interactions (orange) has been explicitly filtered, nuclear 
elastic scattering on hydrogen (yellow), nuclear elastic scattering on oxygen (blue), nuclear elastic scattering on nitrogen 
(magenta), or no nuclear interactions. In all cases, the contribution of Coulomb scattering and ionization remains active.
 
For all considered energies, the dominant contribution on axis throughout the depth of the water phantom stems from 
particle histories where there were no nuclear interactions (green), as expected; this contribution diminishes with 
increasing radial distances from the beam axis. Conversely, the secondaries generated from nuclear reactions of protons 
(purple) have a high contribution at large radial distances and at intermediate depths. Among these secondaries, 
neutrons (and the particles resulting from their nuclear interaction) give rise to the plateau of the purple curves at 
large depths. Interestingly, there are some particular radial distances where the contribution of nuclear elastic 
scattering of protons on hydrogen (yellow), unaltered in FLUKA~\mbox{v4-4.0} with respect to~\mbox{v4-3.4}, becomes 
significant, such as at \mbox{$r=7.62$~cm} for $225$~MeV, and at \mbox{$r=3.05$~cm} and \mbox{$r=4.57$~cm} for 
$160$~MeV. The contribution of nuclear elastic scattering of protons on oxygen (blue) plays an important role both
on axis and out-of-field for all three energies, especially at large depths. Lastly, for $160$~MeV and $100$~MeV, a 
surprisingly important role is played by the nuclear elastic scattering of protons on nitrogen (magenta) in the 
additional air layer in front of the water phantom.

\begin{figure*}
\centering
\includegraphics[width=0.9\textwidth]{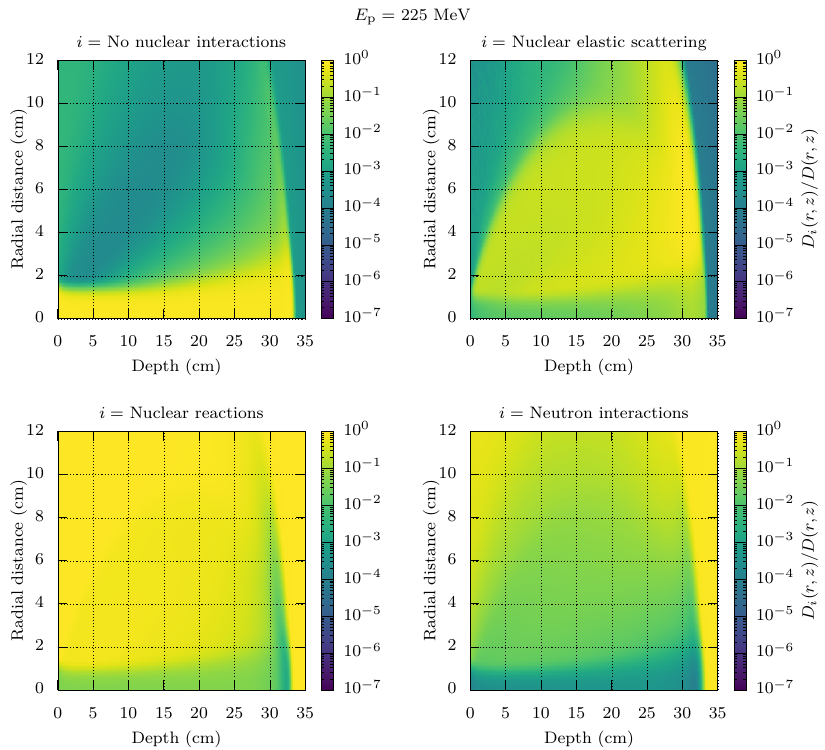}
\caption{Ratio between the FLUKA~\mbox{v4-4.0} absorbed dose due to various interaction mechanisms (see key and text) 
and the total absorbed dose for $225$~MeV protons in water.}
\label{fig:2D_map_225MeV}
\end{figure*}

\begin{figure*}
\centering
\includegraphics[width=0.9\textwidth]{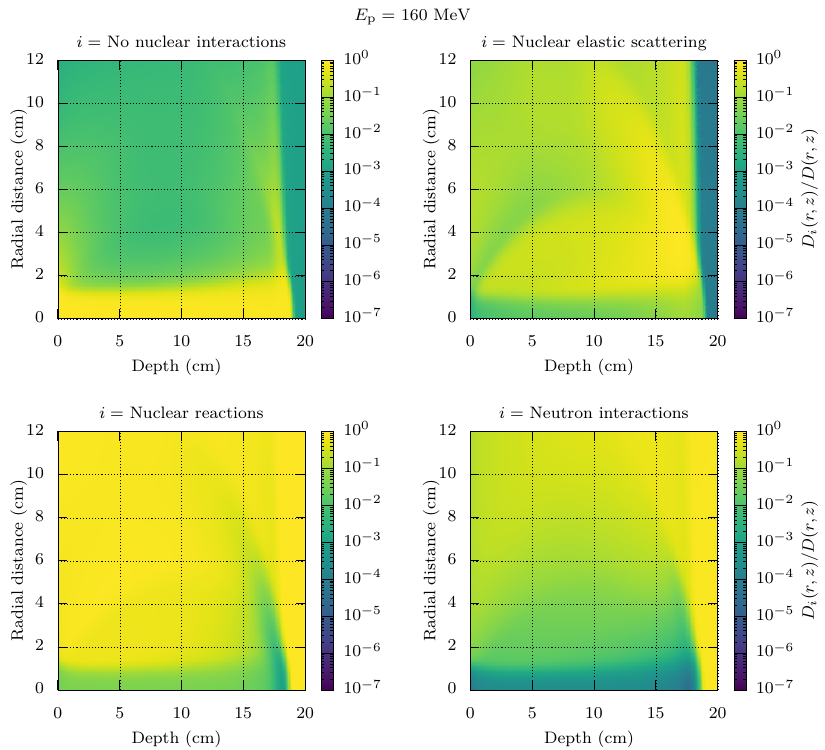}
\caption{Same as Fig.~\ref{fig:2D_map_225MeV} for $160$~MeV protons.}
\label{fig:2D_map_160MeV}
\end{figure*}

\begin{figure*}
\centering
\includegraphics[width=0.9\textwidth]{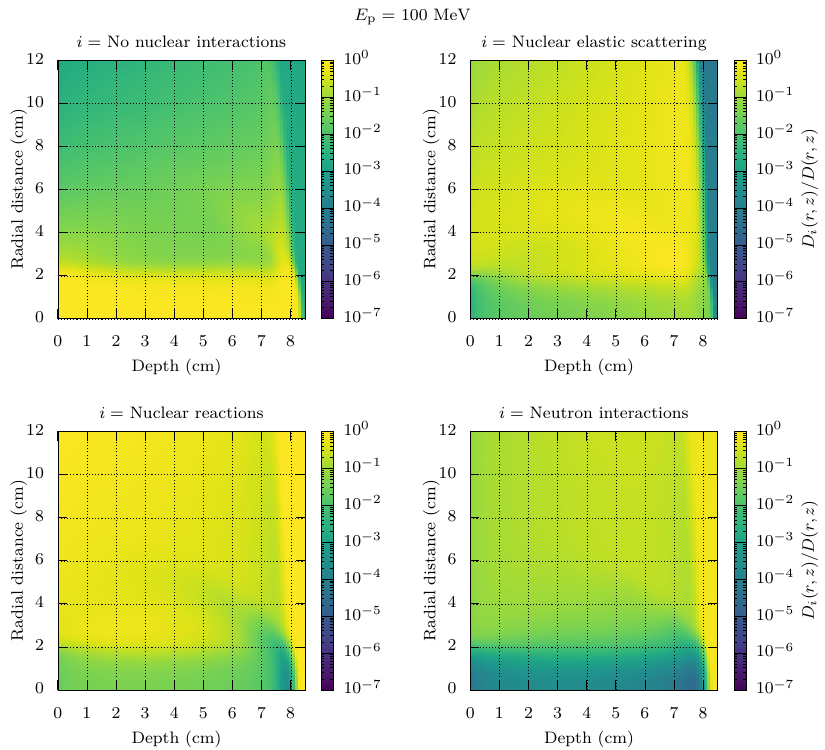}
\caption{Same as Fig.~\ref{fig:2D_map_225MeV} for $100$~MeV protons.}
\label{fig:2D_map_100MeV}
\end{figure*}

To further quantify the role played by each interaction mechanism in the $r$-$z$ dose maps, the ratio $D_i(r,z)/D(r,z)$ has 
been evaluated, where $D_i(r,z)$ is the absorbed dose due to particle histories having undergone an event of kind $i=$\{no 
nuclear interactions, nuclear elastic scattering, nuclear reactions, neutron interactions\} and $D(r,z)$ is the total 
absorbed dose. Figures~\ref{fig:2D_map_225MeV},~\ref{fig:2D_map_160MeV}, and~\ref{fig:2D_map_100MeV} display this ratio as a 
function of both the radial distance from the beam axis and the depth inside the water phantom; $10.5 \times 10^{9}$ primary 
protons have been simulated for these 2D maps. The contribution of proton histories which underwent no nuclear interactions, 
depicted in the upper left panels of the aforementioned figures, exhibits a higher dose ratio near the beam axis, promptly 
diminishing beyond the position of the Bragg peak, around \mbox{$30$-$35$~cm} for $225$~MeV protons, \mbox{$15$-$20$~cm} for 
the $160$~MeV protons, and \mbox{$8$-$9$~cm} for the $100$~MeV protons. As expected, the radial spread (driven by multiple 
Coulomb scattering) of this contribution is relatively narrow, with a significant contribution concentrated close to the 
beam axis that broadens at the end of the proton range. The maximum energy transfer from $100$~MeV, $160$~MeV, and $225$~MeV 
protons to delta rays is of $229.19$~keV, $377.79$~keV, and $548.19$~keV, respectively. At these energies, electrons can 
emit Bremsstrahlung photons, which can travel farther from the beam axis, thus explaining the non-zero dose ratio at large 
radial distances at all depths, even beyond the Bragg peak. Moreover, the particle tracks observed in the upper left panels 
for the $100$~MeV and $160$~MeV proton beam cases are due to secondary electrons emitted by these Bremsstrahlung photons via 
Compton scattering or the photoelectric effect.

The contribution of histories where nuclear elastic scattering occured, displayed in the upper right panels, has a broader 
radial spread than that of histories where no nuclear interactions occured for all three energies, since the differential 
cross section for proton nuclear elastic scattering is broader than that for Coulomb scattering~\cite{serban}. Moreover, 
even if histories where nuclear elastic scattering occured contribute to the overall dose less than those where no nuclear 
interactions occured, this interaction mechanism plays a significant role at greater depths for all radial distances.

The contribution of histories where protons underwent nuclear reactions, depicted in the bottom left panels, extends 
over a wide radial and depth range. Their contribution becomes more significant at greater radial distances compared to 
the contribution of histories where no nuclear interactions or nuclear elastic scattering events occured. This is due to 
the production of secondary particles that can travel farther from the primary beam path. In particular, the 
contribution of histories where neutrons were produced, shown in the bottom right panels, is generally small, but it 
increases at larger depths for all radial distances. In water, neutrons abundantly undergo elastic scattering, which 
decreases their energy until they thermalize. These thermal neutrons have a high capture cross section and lead to the 
emission of $\sim$MeV photons, which can travel $\mathcal{O}(10)$~cm inside the water phantom, populating the large 
depths.

\begin{figure*}
\centering
\includegraphics[width=0.9\textwidth]{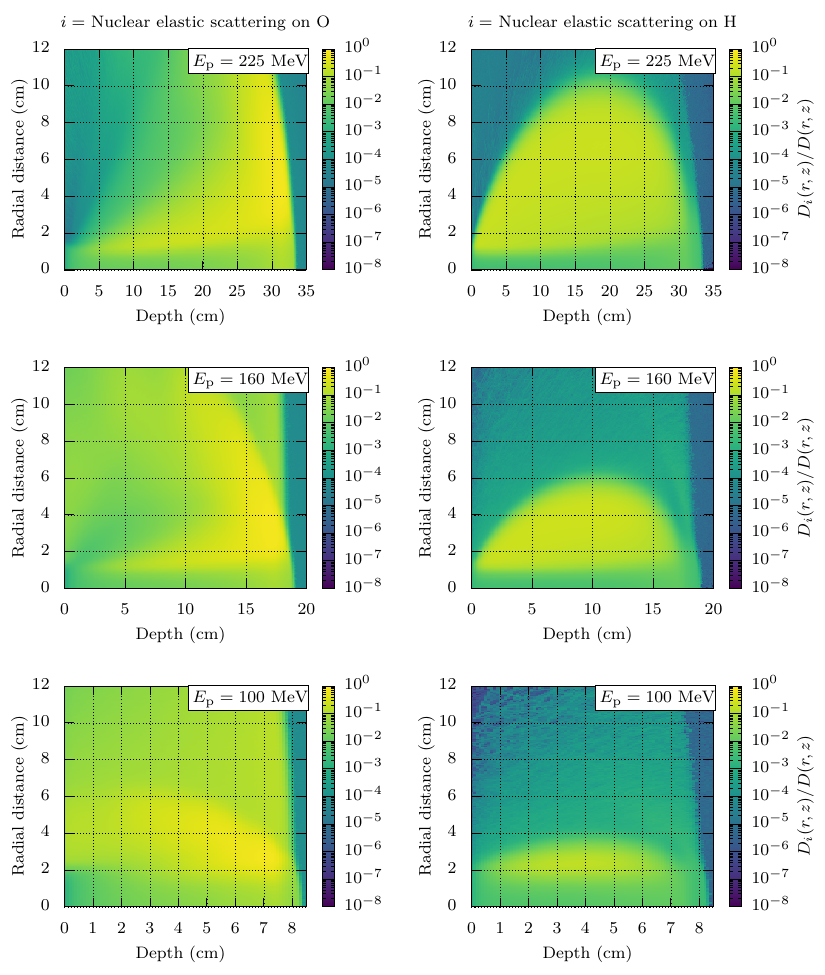}
\caption{Ratio of the FLUKA~\mbox{v4-4.0} absorbed dose due to histories from nuclear elastic scattering of protons on 
oxygen (left panels) and hydrogen (right panels) to the total absorbed dose for $225$~MeV, $160$~MeV, and $100$~MeV protons 
(first, second, and third row, respectively) in water.}
\label{fig:2D_map_ne}
\end{figure*}

The contribution of histories where nuclear elastic scattering of protons occured has been further split according to 
the target nucleus: oxygen or hydrogen. Figure~\ref{fig:2D_map_ne} displays these two contributions on the left and 
right panels, respectively, for the $225$~MeV, $160$~MeV, and $100$~MeV proton beams. In the course of a 
proton-hydrogen nuclear elastic scattering event, due to their equal masses, a larger momentum transfer might occur 
than in proton-oxygen elastic scattering. This large momentum transfer explains the broader radial extent of the 
absorbed dose in the case of a proton nuclear elastically scattered on a hydrogen nucleus, rather than on an oxygen 
nucleus, displayed in the right and left panels of Fig.~\ref{fig:2D_map_ne}, respectively.

\bibliographystyle{elsarticle-num} 
\bibliography{ref}

\biboptions{sort&compress}

\end{document}